\documentclass{iau} 
\usepackage{graphicx}
\usepackage{natbib}

\title[IAUS 359~~ AGN fueling and feedback] 
{Circum-nuclear molecular disks: role in AGN fueling and feedback}

\author[Francoise Combes]   
{Francoise Combes}

\affiliation{Observatoire de Paris, LERMA, Coll\`ege de France, CNRS, \\PSL University, 
Sorbonne University, UPMC, Paris \\ email: {\tt francoise.combes@obspm.fr} }

\pubyear{2020}
\volume{359}  
\jname{Galaxy Evolution and Feedback Across Different Environments}
\editors{T. Storchi-Bergmann et al.}
\begin{document}

\maketitle

\begin{abstract}
Gas inflows fueling AGN are now traceable at
high-resolution with ALMA and NOEMA.
Dynamical mechanisms are essential to exchange angular
momentum and drive the gas to the super-massive black hole.
While at 100pc scale, the gas is sometimes stalled in nuclear
rings, recent observations reaching 10pc scale (50mas),
inside the sphere of influence of the black hole, may bring smoking
gun evidence of fueling, within a randomly oriented nuclear molecular disk.
AGN feedback is also observed, in the form of narrow
and collimated molecular outflows, which point towards the radio mode,
or entrainment by a radio jet. Precession has been observed in a molecular
outflow, indicating the precession of the radio jet. One of the best candidates
for precession is the Bardeen-Petterson effect at small scale, which
exerts a torque on the accreting material, and produces an extended disk warp.
 The misalignment between the inner and large-scale disk, enhances
the coupling of the AGN feedback, since the jet sweeps a large
part of the molecular disk.
\end{abstract}

\firstsection 
\section{Introduction}

 It is now well established that there exists a tight relation between
 the mass of the central supermassive black hole, and the bulge mass,
 or central velocity dispersion, which has been interpreted as
 a co-evolution of galaxies and black holes
  \citep[e.g.,][]{Kormendy2013, Heckman2014}.
This co-evolution might be due to a common feeding mechanism, either
through mergers or cosmic gas accretion followed by secular evolution,
as recenty reviewed by \cite{Storchi2019}, and/or to AGN feedback mechanisms,
regulating the star formation in the galaxy host 
\citep[e.g.,][]{Fabian2012,Morganti2018}.

The new frontier in this domain is to understand in more details
the feeding and feedback mechanisms at the highest possible resolution,
in the complex circumnuclear region, surrounding the black hole, with
the help of multi-wavelength observations
\citep{Ramos2017}. A new view is emerging, where the absorbing material
is not due to the long-expected dusty torus \citep{Hoenig2019}.
VLT Interferometer (VLTI) observations showed
that the dust on parsec scales is not mainly in a thick torus,
but instead in a polar structure, forming like a hollow cone, perpendicular
to a thin disk \citep[e.g.,][]{Asmus2016, Asmus2019}.
 The circumnuclear region, as a transition between the Broad Line Region (BLR)
 of the accretion disk, and the Narrow Line Region (NLR), is complex and
 clumpy, and contains both 
inflowing material in a thin disk, where millimeter lines have been found, 
with also H$_2$O masers, and an 
outflowing component, in the perpendicular direction
\citep[e.g.,][]{Cicone2014, Santi2016}.

In the following, I review ALMA observations at high-angular
resolution of the molecular gas, revealing nuclear trailing
spiral features, that explain the
feeding of the central black hole, through exchange of
angular momentum.
 ALMA observations have also revealed outflows, some
 being extremely collimated in a molecular jet.
 These outflows must be due to the radio mode of AGN feedback,
 even when no radio jet has yet been detected. 
 ALMA has also revealed in most nearby Seyferts the existence
 of molecular circum-nuclear disks, misaligned with the large-scale disks,
 and with decoupled kinematics.
 We identify these parsec-scale structures as
 molecular tori, able to obscure the central accretion disks.
 Several mechanisms are reviewed to explain the misalignments.

\begin{figure}[t]
\begin{center}
 \includegraphics[width=0.99\textwidth]{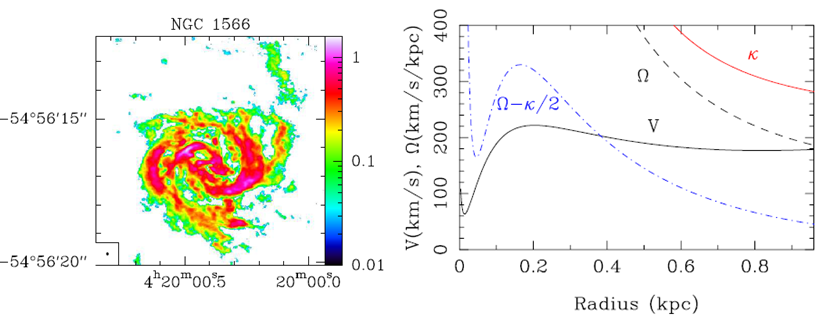} 
 \caption{ALMA observations of the Seyfert-1 galaxy NGC~1566. Left is a zoomed
   8''x8'' region of the CO(3-2) intensity map, showing the nuclear 
	trailing spiral (1''=35pc).
   Right, is the model rotation curve from NIR images, 
	represented schematically,
with the circular velocity (black) epicyclic
frequency $\kappa$ (red), and corresponding $\Omega-\kappa$/2 curve (blue) 
	within the central kpc.
 The contribution of a super-massive black
hole in the nucleus with MBH = 8.3 × 10$^6$ M$_\odot$ has been included.
	From \cite{Combes2014, Combes2019}.}
   \label{fig1}
\end{center}
\end{figure}

\section{Feeding the monster}

To fuel the central black hole, the main problem is to 
transfer the angular momentum of the gas outwards. This can be
done by the gravity torques exerted by bars on the gas in spiral 
arms \citep{Santi2005}. Torques are positive outside corotation, and 
the gas is driven outward to accumulate in a ring at the Outer Lindblad
Resonance (OLR). Inside corotation, torques are negative, and gas
is driven inward, to pile up in a nuclear ring at the Inner Lindblad
resonance (ILR).

What happens inside the ILR depends on the winding sense of orbits there.
The gas is orbiting in elliptic streamlines, which 
gradually tilt by 90$^\circ$ at each resonance 
and wind up in spiral structures. The precession
rate of these elliptical orbits is 
equal to $\Omega-\kappa$/2, with $\Omega$ the rotation frequency = V/r, and
$\kappa$ the epicyclic frequency. 
Usually, inside ILR, and far from the black hole, $\Omega-\kappa$/2
increases with radius, and the spiral is leading. The torque of the bar 
is positive, and the gas is driven back to the ILR. But near the massive black
hole, the precessing frequency $\Omega-\kappa$/2 is decreasing with radius,
and the spiral is trailing. The gas can then fuel the AGN
 \citep{Buta1996}. 

 ALMA has the resolution to
 enter the sphere of influence of the black hole, and 
 a trailing nuclear spiral was
first seen in NGC~1566 \citep{Combes2014}. This nuclear spiral
is located well inside the r=400pc ring, corresponding to the ILR
of the bar.
 Fig. \ref{fig1} shows the nuclear spiral in
 the CO(3-2) line (left), and at right the 
 rotation curve and corresponding frequencies, derived 
 from the stellar potential traced by near-infrared images.
 The precessing rate $\Omega-\kappa$/2 increases towards the center,
 inside 50pc, due to a black hole of mass 8.3 × 10$^6$ M$_\odot$.
 Such trailing nuclear spirals have been found also in NGC~613
 \citep{Audibert2019} and in NGC~1808 (Audibert et al. 2020, in prep.).

\section{AGN feedback: jets and winds}

\begin{figure}[t]
\begin{center}
 \includegraphics[width=0.99\textwidth]{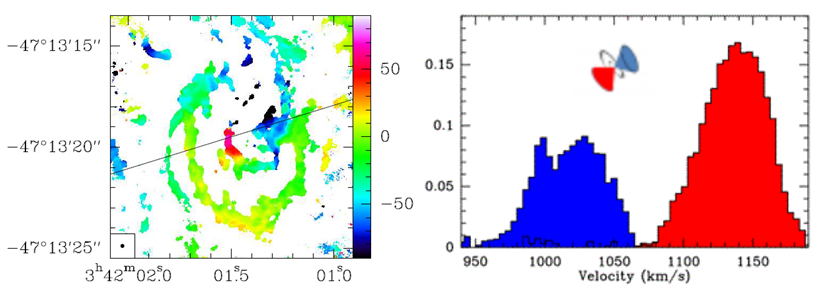} 
 \caption{ALMA CO(3-2) observations of the Seyfert-2 NGC~1433: Left, 
	the velocity field, with the color bar labelled in km/s;
	The thin line indicates the minor axis (PA=109$^\circ$).
	Right, spectrum summing the blue and red-shifted components
	close to the center, along the minor axis. The systemic velocity
   is V$_{sys}$= 1075 km/s. The mass in the outflow is 3.6 10$^6$ M$_\odot$. 
   From \cite{Combes2013, Combes2019}.}
   \label{fig2}
\end{center}
\end{figure}

Molecular outflows are now commonly observed as
AGN feedback \citep{Cicone2014}. If NGC~1566 does not
reveal any outflow, both inflow and outflow can be
observed simultaneously, as in NGC~613,
where a very short (23pc) and small velocity (300km/s)
outflow is detected on the
minor axis, parallel to the VLA radio jet
\citep{Audibert2019}. 
A very small molecular outflow is also seen in NGC~1433,
along the minor axis, cf Fig. \ref{fig2}. This might be the smallest
molecular outflow in a nearby Seyfert galaxy, and could be associated
to a past radio jet 
  \citep[e.g.,][]{Combes2013, Smajic2014}.

\bigskip

In these nearby low-luminosity AGN, which accrete far below
the Eddington limit, the main mechanism
to drive molecular outflows is the radio mode,
i.e. entrainement by the radio jets.
  In some more luminous cases, where L approaches L$_{Edd}$/100,
  there could be both the radio mode, and winds generated
  by radiation pressure (either in the ionized gas, or on dust).
This might be the case of the prototypical Seyfert-2 NGC~1068,
where there is clearly a molecular outflow parallel to the radio jet,
sweeping part of the galactic disk \citep{Santi2014}. The jet
is not perpendicular to the plane, due to the misalignment
of the accretion disk with the plane. The ALMA observations
of the various CO rotational lines reveal clearly a molecular
torus, almost edge-on, and a molecular flow in the perpendicular
direction, aligned with the polar dust. The molecular disk appears
warped and tilted with respect to the
H$_2$O maser disk \citep{Santi2016}.

\bigskip

The lenticular galaxy NGC~1377 is an exceptional case, with
a very thin and highy collimated molecular outflow, in the 
absence of any detected radio jet
 \citep{Aalto2016}. The molecular outflow changes sign
along the flow, on each side of the galaxy. This means that the
jet is almost in the plane of the sky, and that a slight precession  
of only  10$^\circ$ is able to tip the jet from redshifted to 
blue-shifted and back. Such a precession is
observed in micro-quasars jets in the Milky Way, for instance SS433
\citep{Mioduszewski2005}. But this can be attributed to the companion
star. Here there must exist another origin of the precesion, which 
could be relativistic (see next section).

A precessing molecular outflow model is compatible 
with the data \citep{Aalto2016}.
The flow is launched close to the center (r $<$ 10pc).
A radio jet must exist at a low level, or has existed in a recent past.

\section{Molecular tori: misalignment}

With the high spatial resolution of ALMA, it was possible to
unveil circumnuclear disks in the CO emission, towards nearby
Seyferts. These happen to be misaligned to the large-scale disks,
and kinematically decoupled. We call them molecular tori, 
they exist in 7 out of the 8 cases observed \citep{Combes2019}.
The average radius of the molecular tori is 18~pc, with a median
at 21~pc. Their average molecular mass is M(H$_2$) = 1.4 10$^7$ M$_\odot$,
and in average their inclination on the plane of the sky is
29$^\circ$ different from their galactic disk.

These molecular tori are clearly within the sphere
of influence of their black holes, and can serve to measure their mass,
provided that their inclination is sufficient
 \citep{Combes2019}. This has been done also for more massive
early-type galaxies,  by the WISDOM project
\citep{Davis2018}.

\bigskip
We can invoke at least three mechanisms of misalignment
between the large-scale galactic disks and the 
molecular tori and/or accretion disks.
One of them is the
radiation-driven warping instability \citep{Pringle1996}.
A tilted optically thick disk, which absorbs the
radiation from the central AGN, receives in each point
some momentum from the radiation, but no torque, 
because of the radial direction.
But then it re-radiates perpendicularly to its orientation, and this
produce torques, which maintain and amplify the warping.
Assuming the luminosity is powered by accretion eliminates
the unknown viscosity parameter $\alpha$. The instability 
occurs for radii R $>$ 0.1pc M$_{BH}$/(10$^8$M$_\odot$). The
efficiency of the mechanism was tested by simulations, both
in the case of retrograde and prograde precession with respect
to the disk rotation \citep{Maloney1997}.
A second meschanism is the magnetic instability, and consequent torques
but compatibility with AGN observations is contrived, it is more 
adapted to accretion disks around magnetic stars
\citep{Pfeiffer2004}.

\begin{figure}[t]
\begin{center}
 \includegraphics[width=0.99\textwidth]{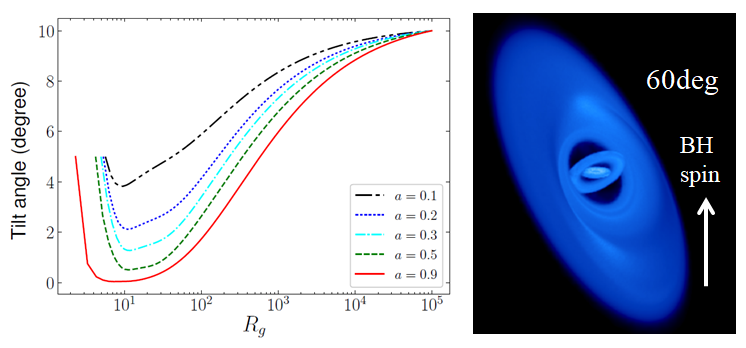} 
 \caption{Misalignment of the accreting material near a black hole:
Left: radial profiles of the disk tilt-angle (with respect to 
the black hole spin), for several values
of the parameter a, dimensionless angular momentum $J$ of the black
hole $M$, 0 $<$ a $<$ 1 (a= c$J$/(G$M^2$)). The initial tilt-angle is 
5$^\circ$. The radius is in unit of the gravitational radius 
	R$_g$ = G$M$/c$^2$. From \cite{Banerjee2019a}. 
Right: Simulation of the Bardeen-Petterson effect, in a disk initially
inclined by 60$^\circ$ with respect to the black hole spin. The Lense-Thirring
precession causes the disk to break in a few tilted rings.
	Adapted from \cite{Nealon2015}.}
   \label{fig3}
\end{center}
\end{figure}

A third mechanism uses the
Bardeen-Petterson effect \citep{Bardeen1975}, due to 
Lense-Thirring precession. The accretion disk has a random orientation,
generally not aligned with the black hole spin. The relativistic
frame dragging effect induces a precession, which tends to align
the inner parts of the accretion disk with the black hole equator.
The disk develops a warping up to distances 10$^2$ to 10$^4$ 
Schwarzschild radius R$_s$. 
The precession of the disk and its warp can be seen
from inner to outer disk, up to 1~pc M$_{BH}$/(10$^9$M$_\odot$).
According to the amplitude of viscosity, one can distinguish
two regimes: the diffusion, when $\alpha >$ H/R, where H is the height 
of the disk, and the regime of bending waves, when  $\alpha <$ H/R
\citep{Papaloizou1983}. In the first case, the disk is warping smoothly
and continuously, while in the second case, the disk can break in several rings,
with different inclinations and precessing rates. Then the various rings,
with differential precession, collide, and drive the gas to fuel the AGN
more quickly. This regime has been simulated by \cite{Nealon2015}.
 Some works found that the alignment of accretion disks with
 the black hole equator, through the Bardeen-Petterson effect,
 was inefficient \citep{Zhuravlev2014, Banerjee2019a, Banerjee2019b}.
 More precisely, according to some viscous parameters (parallel of 
 perpendicular to the disk), 
 and viscosity generated by magnetized turbulence, the disk near the 
 black hole can retain its inital inclination, instead of aligning, 
 cf. Fig. \ref{fig3}. Then precessing jets can be launched, perpendicular to the
 disk, but not aligned to the black hole spin
  \citep{Liska2018,Liska2019}.

A manifestation of these warping instabilities is
the observation of the warped maser disks.
Water masers in the prototypes NGC~4258 \citep{Herrnstein1999} and NGC~1068
\citep{Gallimore2004} are detected on 0.3-0.8~pc warped discs. 
These observations
are best represented by the Lense-Thirring effect
and/or the radiation driven warps. These perturbations
also heat the disk. The fitting of the observations has been
done for NGC~4258 by \cite{Martin2008}, and for NGC~1068 by \cite{Caproni2006}.
In NGC~1068, where the Bardeen-Petterson mechanism gives the best
fit with the observations, the disk is aligned with the black hole spin,
until the radius R$_{BP}$ = 10$^{-5}$ to 10$^{-4}$ pc, then warps. 
For one of the best fit models the
alignment time-scale is 7580 yr, the misalignment angle
 40$^\circ$ and the velocity of the jet 0.17c. The shape and precession
of the pc and kpc-scale jet is also fitted, 
following \cite{Wilson1987}, in addition
to the warped H$_2$O maser disk by \cite{Gallimore2004}.

 How can the gas accreted from the galactic disk be so 
 misaligned with the disk itself?
 First, the potential in the center is almost spherical, and the disk
 very thick with respect to the parsec-scales in question, and second,
  star formation feedback constantly ejects some gas out of the plane,
  which rains down in a fountain effect at a random orientation,
sometimes in a polar ring \citep{Renaud2015, Emsellem2015}.

\section{Summary}

Thanks to the high resolution provided by ALMA on the molecular gas,
it is now possible to better understand the AGN fueling mechanisms.
If at large scale, the primary bars can drive the gas towards the 100pc
scales, in ILR rings, the nuclear bars act on a trailing nuclear
spiral to drive the gas towards the black hole, when the circumnuclear
gas enters its sphere of influence. Inside the nuclear spiral,
ALMA has revealed the existence of morphologically 
and kinematically decoupled circum-nuclear disks,
or molecular tori.

In some less frequent cases, we can see both AGN fueling,
and molecular outflows, through AGN feedback. This can occur via
entrainment by radio jets, or through disk winds (or both).
 The radio mode is distinguished by extremely thin and collimated
 molecular jets, sometimes precessing, as are the entraining radio jets.

To explain the precession, and the misalignment of the circum-nuclear disks,
we can invoke at small scale the Bardeen-Petterson effect, which produces
torques in the accreting material, to align it with the black hole spin, and
then induces a warping up to a fraction of parsec-scale. 
The black hole is likely to be fueled via several accretion episodes, 
coming from the large-scale galacic disk, but also the fountain gas, 
ejected above the plane through supernovae feedback.

\end{document}